# Optical design trade-off study for the AO module of MAVIS

Greggio D.[a,f], Di Filippo S. [a,f,h], Magrin D. [a,f], Schwab C.[b], Viotto V. [a,f], Busoni L.[c,f], Esposito S. [c,f], Ragazzoni R. [a,f], Fusco T.[d], Benoit N.[d], Pinna E. [c,f], Rigaut F.[e], Arcidiacono C. [a,f], Bergomi M. [a,f], Biondi F. [a,f], Chinellato S. [f,g], Farinato J. [a,f], Marafatto L. [a,f], Portaluri E. [a,f], Santhakumari K.K.R. [a,f], Vassallo D. [a,f]

[a]INAF Padova, Vicolo dell'osservatorio 5, 35122, Padova, Italy;
[b]AAO-MQ, Macquarie University, Sidney, Australia;
[c] INAF Arcetri, Largo Enrico Fermi, 5, 50125 Firenze, Italy;
[d]Laboratorie d'Astrophysique de Marseille, 38 Rue Frédéric Joliot Curie, 13013 Marseille, France
[e]AAO - Stromlo, RSAA, Australian National University, Canberra ACT 0200, Australia
[f]ADONI, Laboratorio Nazionale di Ottica Adattiva Italiano
[g] INAF-Direzione Scientifica, V.le del Parco Mellini 84, 00136 Roma, Italy;
[h] Università di Padova, Dipartimento di Fisica e Astronomia, Vicolo dell'osservatorio 3, 35122, Padova, Italy.

## ABSTRACT

MAVIS (MCAO-Assisted Visible Imager and Spectrograph) is an instrument proposed for the VLT Adaptive Optics Facility (AOF), which is currently in the phase-A conceptual design study. It will be the first instrument performing Multi-conjugate adaptive optics at visible wavelengths, enabling a new set of science observations. MAVIS will be installed at the Nasmyth platform of VLT UT-4 taking advantage of the already operational Adaptive Optics Facility that consists of 4 LGS and an adaptive secondary mirror with 1170 actuators. In addition, two post-focal deformable mirrors and 3 Natural Guide Stars (NGS) are foreseen for the tomographic reconstruction and correction of atmospheric turbulence.
The MAVIS AO module is intended to feed both an imager and a spectrograph that will take advantage of the increased resolution and depth with respect to current instrumentation.
In this paper we present the trade-off study for the optical design of the MAVIS AO module, highlighting the peculiarities of the system and the requirements imposed by AO. We propose a set of possible optical solutions able to provide a compact and efficient implementation of the different subsystems and we compare them in terms of delivered optical quality, overall throughput, encumbrance, ease of alignment and residual distortion.
**Keywords:** MAVIS, Optical design, relay optics, design optimization, MCAO, visible adaptive optics, VLT

## 1. INTRODUCTION

MAVIS is the VLT MCAO-Assisted Visible Imager and Spectrograph, which will provide AO corrected imaging and spectroscopic capabilities at visible wavelengths over a field of view of 30x30 arcsec. MAVIS will be installed at the VLT UT4 Nasmyth platform and will use the already operational Adaptive Optics Facility (AOF), currently consisting of 4 LGS and an adaptive secondary mirror with 1170 actuators [1]. The design of MAVIS is based on a modular approach, in which each module performs its own specific tasks minimizing the interactions with the other sub-modules as much as possible. This paper describes the optical design trade-off study of the Adaptive Optics Module (AOM) of MAVIS, which is a self-contained MCAO module, in the visible, delivering a corrected FoV to the post-focal scientific instruments, comprised of the Imager, Spectrograph and an additional port for a third instrument. Following the modular approach, also the AOM is composed by different sub-modules having the following functionalities:

- **Post Focal Relays module (PFR):**
    - receive light from VLT
    - re-image meta-pupils onto 2 post-focal DMs (at predetermined altitudes)
    - feed the NGS WFS with a 2 arcmin FoV at infinity (NIR band)
    - feed the LGS WFS with a 1 arcmin FoV at 90-230km altitude
    - deliver the MCAO-corrected 30" diameter FoV to the Imager and Spectrograph (VIS band)
    - provide at least 2 output ports and include a means to switch between them
    - provide field de-rotation for the Instruments and NGS WFS, by means of a k-mirror
    - correct for atmospheric dispersion
- **NGS WFS**: it is the Low Order (LO) wavefront sensor of MAVIS. The current baseline assumes 3 NGSes can be sensed, in the J+H bands, at the same time, in a 2 arcmin diameter FoV.

- **LGS WFS:** it is the High Order (HO) wavefront sensor of MAVIS. The current baseline assumes 8 LGSes can be sensed at the same time, with a circular asterism.

In this paper, we present and discuss possible optical configurations for the PFR module, which is the core optical system of MAVIS, providing the interface towards the instruments and all the other sub-modules. More details on the system architecture can be found in the paper by Francois et al. (2019, this conference) [2].

## 2. DESIGN REQUIREMENTS

The design requirements used during the phase A study were derived by a combination of top-level requirements and some assumptions based on experience. This was necessary in order to start developing the optical design while other trade-off studies were running in parallel. We describe below the most important requirements and their implications on the optical design. Some of them shall be intended as soft requirements used to guide the design trade-off rather than hard requirements that preclude the functioning of the instrument.

**Modularity**

Modularity is fundamental to ease integration and testing, but means that the PFR optics shall deliver very clear and possibly testable optical interfaces. This will help checking that all interfaces are within specifications during the alignment and integration of every sub-module and is also valuable for maintenance activities.
From the optical design point of view, the goal is to deliver well-corrected and accessible focal planes to every sub-module, namely the LGS and NGS WFSs and the instruments output ports.

**De-rotation scheme**

The preferred de-rotation scheme foresees an optical de-rotator at the entrance of the instrument to compensate for sky field rotation. An additional de-rotator is necessary for LGS tracking, because the AOF LGS asterism is fixed with respect to the telescope pupil. LGS de-rotation can be either optical or mechanical.

**MCAO related requirements**

In order to achieve wide-field correction, we require two additional post-focal DMs conjugated at higher altitudes with respect to the VLT entrance pupil. The DMs are part of the PFR and they need to be placed upstream the NGS and LGS WFSs to allow closed-loop operation. Preliminary requirements from AO simulations indicate a projected pitch size around 25-30 cm and conjugation altitudes of 4km and 12km. These values have been taken as a reference rather than as strict requirements during design optimization.

For the post-focal DMs we identified the ALPAO DM3228 (1.5mm actuator pitch) as a possible solution because it presents a good compromise between overall meta-pupil size (93mm) and the number of actuators (64 over the diameter). For this reason, we assumed a pitch size of 1.5mm and meta-pupil size in the range 80-90mm during the trade-off study. Having meta-pupils much smaller or much larger than this, by using different DMs technologies, will lead either to a poor optical quality (in the case of small pupils) or to increased volume and size of optical components (in the case of larger pupils).

**Number of output ports**

At least two output ports are foreseen, with a goal of a third output port for visiting instruments. This requirement involves the accessibility of the output focal planes to the science instruments and the allocation of volume around them along with a mechanism for output port selection.

**Waveband**

MAVIS is designed to work down to the V band, corresponding to a waveband from 450 nm to 950 nm for the science instruments, while tip-tilt wavefront sensing is done in the NIR, where the waveband 950-1700nm has been chosen in order to avoid the need for a fully cryogenic WFS.

Concerning the science spectral coverage, there is interest to push down the blue limit up to 370nm in order to include some relevant spectral features. This of course has an impact on the optical design, especially in the case of refractive components. For the purposes of phase A design trade-offs, we decided to select glasses with reasonable throughput down to 370nm, but optimization is restricted to the 450-950nm wavelength range in order to avoid performance degradation at V band.

**Field of view**

The major requirement on the field of view is set by sky coverage reasons, for which a technical 2 arcmin diameter FoV is necessary for NGS selection. For LGS the design has been optimized over a 1 rcmin diameter FoV in order to allow enough freedom for the choice of the best asterism. The field of view delivered to the instruments is 30x30 arcsec.

**Optical interface requirements**

The optical interface towards the other sub-modules in terms of plate scale is not strictly fixed at this stage of the design. However other parameters are important for operational and calibration reasons. For example, it is desirable to have a telecentric NGS focal plane in order to avoid compensating for chief ray tilt when selecting different reference stars across the field.

Concerning optical interface towards the LGS WFS, it is desirable to have a fixed exit pupil position when focusing at different heights (this can be achieved, for instance, by having a telecentric output beam, and mechanically refocusing by moving the whole LGS WFS assembly).

Another important parameter is distortion. The minimization of distortion during design optimization is fundamental both to simplify calibration procedures and to maximize the astrometric performances of the instrument. Distortion shall be minimized at the science focal plane as well as at the NGS focal plane, to reduce TT loop NCPAs.

Finally, field curvature shall also be minimized, particularly at the level of NGS focal plane, to avoid the need for refocusing when selecting a reference star across the field. Some residual field curvature can be accepted at the LGS and science focal planes.

**Design parameters summary**

We report in Table 1 a summary of the main design parameters adopted for the trade-off study.

| DMs | |
|---|---|
| DM_High conjugation height [km] | 12 |
| DM_Low conjugation height [km] | 4 |
| DM_High projected pitch [cm] | ~30 |
| DM_Low projected pitch [cm] | ~20 |
| **NGS sub-module** | |
| Radial field of view [arcsec] | 60 |
| Waveband [nm] | 950-1700 |
| Telecentricity | Telecentric (req. TBD) |
| **LGS sub-module** | |
| Radial field of view [arcsec] | 30 |
| Conjugation distance [km] | 90-230 km |
| Telecentricity | Telecentric (req. TBD) |
| **Science instruments** | |
| Radial field of view [arcsec] | 21.2 |
| Optimization waveband [nm] | 450-950 |

*Table 1: AOM optical requirements and assumptions.*

## 3. INVESTIGATED OPTICAL CONFIGURATIONS

There are three main classes of solutions studied for the AOM optical design:

1. **Refractive designs** based on on-axis optical elements. The general advantages of this class of solutions are rotational symmetry, manufacturability and ease of alignment. The main disadvantages are chromatic aberrations and lower throughput in parts of the wavelength range (driven by the choice of glasses transparent in the 370-1700nm waveband).
2. **Reflective designs** based on off-axis mirrors. The main advantage is achromaticity, but the resulting system is not rotationally symmetric and generally more difficult to align and manufacture. Moreover, in order to compensate aberrations and distortion, some sort of symmetry in the position of optical elements should generally be exploited, leading to geometrical constraints and less design flexibility.
3. **Catadioptric designs:** a combination of the two above. A reasonable choice is to use off-axis mirrors in the common path, before splitting the light towards the NGS and LGS sensors. In this way, chromatic aberrations are produced in the single sub-modules and can be corrected more easily due to the smaller waveband (450-950nm for the science channel and 950-1700nm for the NGS channel). However, the system is not rotationally symmetric.

For each class of solutions, we developed an optical design as much as possible compatible with the requirements described in the previous section. The three designs are described below.

**Refractive design**

The nominal refractive design is depicted in Figure 1below. The first element is a collimating doublet which forms an image of the telescope pupil of approximately 42mm diameter, close to which an atmospheric dispersion corrector (ADC) consisting of two counter-rotating prisms is placed. Following that, another doublet is used to compress the beam diameter in order to reduce the size of the K-mirror placed immediately after it; it also creates an intermediate focal plane. Before the intermediate focal plane, a field lens generates an image of the meta-pupils (4km and 12km layers) onto the DMs. A dichroic mirror placed in a pupil plane transmits the NIR light to the NGS WFS module and reflects the VIS light, which is then split by a notch filter transmitting the LGS light and reflecting the rest of the visible spectrum towards the instrument ports.

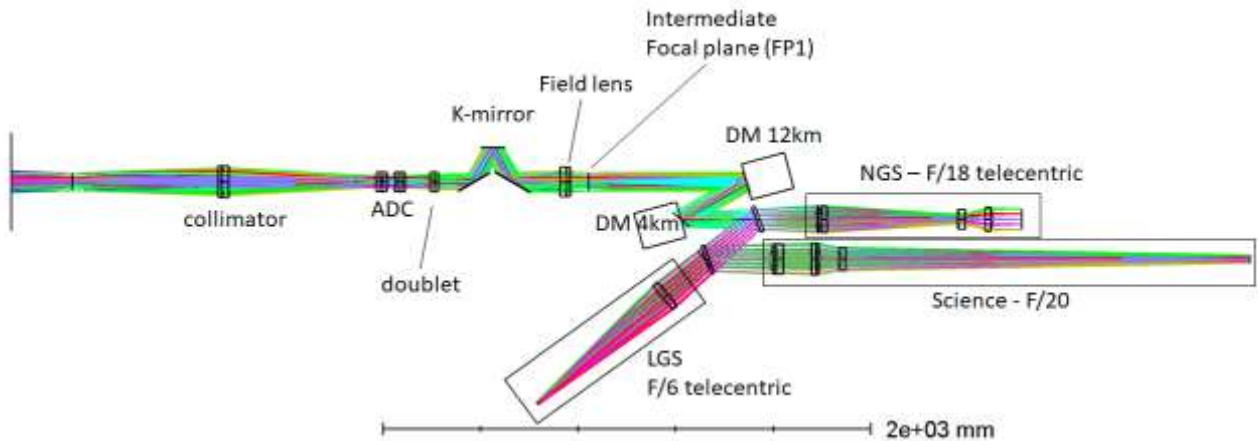

*Figure 1: optical layout of the refractive configuration*

A 4-lens objective on the NGS channel generates a F/18 telecentric and flat focal plane for the NGS sensors.

The LGS focal plane is F/6 and is created by a single aspheric lens optimized over 90-230km conjugation distance. The focusing at different distances is achieved by moving the LGS sensors along the optical axis.

The SCI focal plane is formed by a 5-lens objective delivering an F/20 beam with long focal extraction.

**Reflective design**

For the reflective configuration, different classes of optical relays have been attempted: off-axis parabolic mirrors, Offner configurations, and a hybrid system made of spherical + freeform mirrors. A common challenge of the design of off-axis solutions for the AOM is the need to guarantee the volume for all the required components (DMs, ADC, K-mirror, calibration unit) while using symmetries to cancel out the system aberrations.

An Offner relay which generates beam footprints compatible with the ALPAO DM3228 (1.5mm actuator pitch) does not offer accessibility to two post-focal DMs, even when using a double Offner relay. For this reason, it has been discarded.

The more general and unconventional design based on spherical + freeform mirrors has also been abandoned due to the difficulty to control optimization for all the relevant conjugation planes (meta-pupils, LGS and Sky). In general, the use of freeform surfaces can be very powerful, but it is difficult to control the many degrees of freedom available during optimization. For this reason, even if we cannot discard the possibility that this kind of design for the AOM offers a workable solution, we believe that the level of complexity it introduces is not appealing, also considering the relative lack of expertise with such a design, and experience related to the use of general freeform optics.

The off-axis parabolic mirror design is more promising with respect to the other two all-reflective configurations and is described below. We use a 4-OAP configuration in order to compensate the field distortion generated by the first couple of parabolic mirrors. Moreover, the light is split to the NGS and LGS WFS before the last parabolic mirror. This means that the output focal plane of each channel is formed by a different parabolic mirror, each optimized separately. This is particularly relevant for the LGS channel, where the different conjugation height imposes different geometrical constraints to compensate for off-axis aberrations.

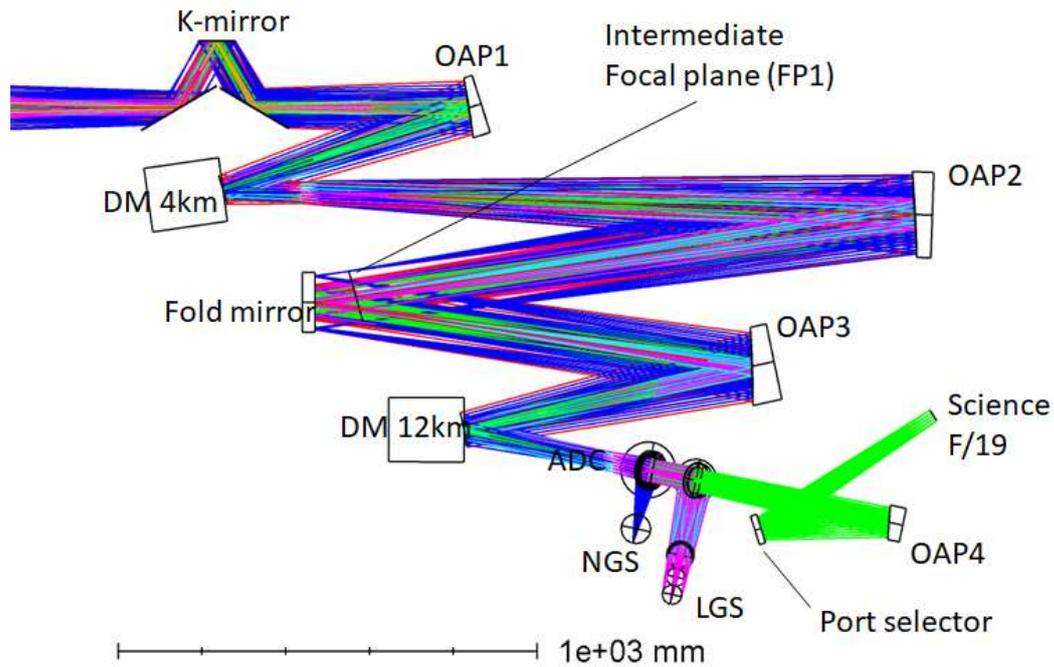

*Figure 2: optical layout of the reflective configuration*

As shown in Figure 2, the first off-axis parabola (OAP1) collimates the light from the telescope, projecting an image of the 4 km atmospheric layer onto the first DM. The second and third parabolic mirrors, together with a flat folding mirror, relay the light to the second DM, conjugated at 12 km altitude. The ADC is placed after that, followed by the NGS dichroic filter and the LGS notch filter. The dichroics are placed in the collimated beam to avoid the generation of astigmatism in the transmitted beams.

The OAP mirror on the NGS channel delivers an F/15 telecentric and curved focal plane with a convex (vertex towards the incoming beam) radius of curvature of 250mm. The OAP mirror on the LGS channel, optimized for a conjugation altitude from 90 km to 230 km, produces an F/12 telecentric and curved focal plane with a radius of curvature of 220mm.

The OAP mirror on the science channel generates an F/19 curved focal plane with a radius of curvature of 315mm.

**Catadioptric design**

The catadioptric design, shown in Figure 3, uses three off-axis conical mirrors in the common path to relay the light onto the DMs and to produce an output collimated beam. The K-mirror and the ADC are also placed in the common path.

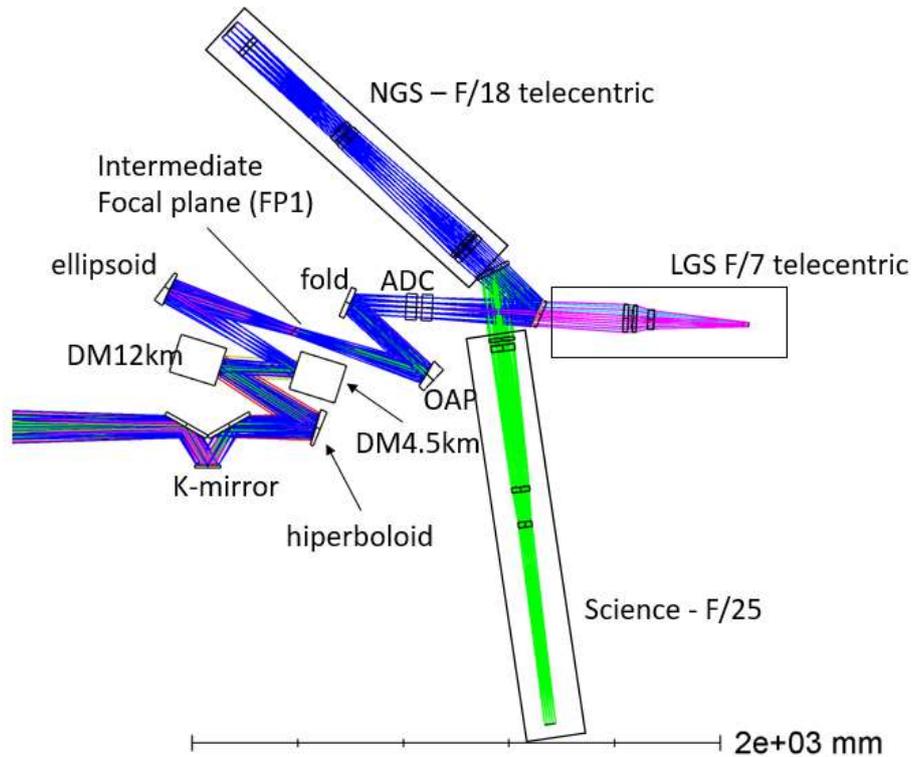

*Figure 3: optical layout of the catadioptric configuration*

The light is then split by a dichroic mirror and a notch filter towards the WFS sub-modules and the science instruments. The focal plane for each module is created by refractive objectives. We stress here that the use of on-axis centro-symmetric optical elements is not optimal for the LGS channel, because the LGS image is affected by field variable aberrations (mainly astigmatism) generated by the three off-axis mirrors and the finite distance of the LGS. The alternative of using off-axis mirrors on the LGS channel has been attempted but is not effective due to the variable conjugation range, which requires more than two off-axis mirrors leading to impractical implementation of the setup. The consequence of this choice is some residual aberrations on the LGS WFSs that need to be compensated or characterized in some way.

## 4. TRADE-OFF CRITERIA

We compared the three designs presented above based on a set of criteria, assigning a score to each of them in order to identify the best trade-off. Below we report the list of criteria used and their relative weight (in brackets):

- Optical performance (5):
    - Image quality [rms WFE] (4)
    - NCPA (4)
    - Distortion (3)
    - Image plane curvature radius (3): min allowed 250mm (instruments req.)
    - Telecentricity (2): max allowed 1deg (instruments req.)
    - Throughput (4)
    - Meta-pupils quality: to be compared with 1.5mm DMs pitch (3)
- Engineering tasks (4):
    - MAIT: manufacturability of optical elements (2)
    - MAIT: accessibility of intermediate focal planes/pupil planes (3)
    - MAIT: ROM alignment tolerances (4)
    - MAIT: number of aspherical surfaces (2)
    - Ease to accommodate calibration unit (3)
    - Volume allocation (3)

- Management/Risks (3)

# 5. DESIGNS COMPARISON

Without entering in the detailed performance of each design (which can be found in Table 2 and Table 3), we discuss below their general strengths and weaknesses.

Concerning optical quality the best configuration is the reflective one. The absence of chromatic aberrations and exploitation of symmetries allows for an almost perfect nominal optical quality (~10nm RMS WFE against a few tens of nanometers of the other two designs), The same is not true for the meta-pupils optical quality at the DMs. Here, the best option is the refractive one, because the other two designs are affected by non-negligible off-axis aberrations (mainly coma). Telecentricity and distortion are quite similar for the three configurations, while field curvature is stronger on the off-axis designs.

Concerning throughput, the best design is the reflective one, due to the lower number of optical surfaces. The main losses are due to two main reasons:
1) The large waveband of the common-path optics (goal 370-1700 nm) requires complex multi-layer coatings which are difficult to manufacture and have ~2% throughput loss for every surface
2) The internal transmittance of glasses in the blue region (370-450nm) is generally lower.

This can be mitigated by reducing the number of optical components on the common path (for instance the K-mirror or the ADC). However, this implies also an increase of NCPAs and differential image motions among the sub-modules with a decrease of performance and increase of calibration complexity.

When taking into account also engineering tasks, the preferred option is definitely the refractive design. This is because on-axis components are easier to manufacture, measure and align. This has also an impact on the final image quality achievable for the real system. In fact, despite the design based on off-axis parabolic mirrors has almost perfect image quality, this is quickly lost when accounting for tolerances, which become the dominant term in the overall error budget. In terms of volume allocation, all the three designs have similar volume, but the distribution of volume is different. The catadioptric design is the most compact, while the refractive design is slightly longer than the Nasmyth platform and requires folding. Another important aspect is the possibility to place the instrument calibration unit (CU) at the input Nasmyth focal plane in order to allow calibration of the whole optical path. Both the off-axis configurations are not completely compatible with this requirement, unless the K-mirror is moved downstream.

Concerning management, cost and risks, all the configurations are similar. The three designs do not show appreciable differences in terms of cost. In general, off-axis optics metrology could have a minor impact on cost budget. This pertains to both, the Reflective and Catadioptric designs. Maintenance activities are expected to be comparable for the three designs. The main risk identified is related to ageing of the coatings, which, for MAVIS, need to show good performance in a wide waveband range. This pushes the selection toward thick multi-layers coatings, which may include residual stresses between layers, that can evolve during ageing. This is common to all of the considered designs.

To conclude, the comparison indicates the refractive design as the preferred alternative. Moreover, some preliminary analyses indicate that:

- Further optimization of glasses selection can lead to an increase in the throughput of the refractive design (the weakest point of this alternative).

- the lack of space for CU accommodation at entrance focal plane, together with the difficulties in the K-mirror positioning are confirmed to be a show-stopper in the assumed de-rotation scheme for the reflective design

- The refractive design is more robust to changes of the design parameters (i.e. DMs sampling and conjugation altitude, optical interface towards sub-modules) and can be re-optimized more easily than the off-axis configurations.

|  | **Refractive** | **Reflective** | **Catadioptric** |
|---|---|---|---|
| **Image quality SCI[1]** | 28-45 nm RMS | 7 nm RMS | 34-39 nm RMS |
| **Image quality LGS[2]** | 17-53 nm RMS | 13-48 nm RMS | 42-200 nm RMS |
| **Image quality NGS[3]** | 36-117 nm RMS | 2-15 nm RMS | 16-41 nm RMS |
| **NCPA[4]** | 50nm | 41nm | 60-70nm (based on extrapolation from the other two designs) |

| Distortion | SCI 0.002%<br>NGS 0.01% | SCI 0.011%<br>NGS 0.033% | SCI 0.06%<br>NGS 0.09% |
|---|---|---|---|
| FoV curvature | SCI: 460mm<br>NGS: 2e4mm | SCI: 315mm<br>NGS: 250mm | SCI: 280mm<br>NGS: 220mm |
| Non-telecentricity angle[5] | SCI: 0.37deg<br>NGS: 0.12deg | SCI: 0.19deg<br>NGS: 0.09deg | SCI: 0.7deg<br>NGS: 0.07deg |
| Throughput | [450-950]nm: 0.58<br>[370-450]nm: 0.31 | [450-950]nm: 0.69<br>[370-450]nm: 0.50 | [450-950]nm: 0.65<br>[370-450]nm: 0.42 |
| Meta-pupils quality [90% EE radius] | DM@4km: 85μm<br>DM@12km: 65μm | DM@4km: 100μm<br>DM@12km: 150μm | DM@4km: 330μm<br>DM@12km: 150μm |

**Table 2 : optical quality design comparison**

**Comments to Table 2:**
[1]Reported values refer to the worst rms WFE within the 30"x30" FoV (actually, the values always correspond to the corner of the FoV). The ranges are min. and max. rms WFE values in the following wavebands: 450-600nm, 560-715nm, 715-950nm.
[2]Reported values refer to the best and worst rms WFE @589nm in the following test cases: 1) off-axis distance 17.5" and 30"; 2) conjugation distance 90km and 230km.
[3]Reported values refer to the best and worst rms WFE within the 1' radius FoV, estimated in the NGS WFS waveband 950-1700nm.
[4]Reported values are the total rms NCPA, due to the fact that the focal planes delivered to the LGS WFS and the instruments are different by design (e.g. in some design LGS WFS focal plane is affected by astigmatism. No alignment, flexures, etc included here). Additionally, the focal plane delivered to LGS WFS is derotated (to follow telescope elevation) in a differential way with respect to the focal plane delivered to the instruments (following the sky). We are here considering all these contributions as NCPA. If we distinguish between a static component (easier to calibrate) and a dynamic one (more complicated to calibrate and related to differential rotation of SCI and LGS FoV), the latter is dominant in the Reflective design. Moreover, such a dominant component is strongly asymmetric, so requiring a better calibration.
[5]The designs are not optimized for this particular parameter. For this reason the weight is kept low.

|  | **Refractive** | **Reflective** | **Catadioptric** |
|---|---|---|---|
| **Manufacturability**[1] | No show-stoppers. | No show-stoppers.<br>Off-axis optics (optical quality + metrology) | No show-stoppers.<br>Off-axis optics (optical quality + metrology) |
| **Intermediate planes accessibility** | 1 entrance FP<br>1 intermediate FP | 1 intermediate FP (changing folding) | 1 intermediate FP<br>1 intermediate PP |
| **Alignment tolerances**[2] | No show-stoppers.<br>focusing: ~0.5mm<br>centering: ~0.5mm<br>tilt: ~0.05deg | No show-stoppers.<br>focusing: ~0.1mm<br>centering: ~0.1mm<br>tilt: ~0.005deg | No estimation done. Assumed to be similar to reflective design. |
| **Number of aspherical surfaces** | 1 surface. Small aspheric departure. | 0 surfaces | 0 surfaces |
| **Calibration Unit accommodation** | CU deployable mirror (or similar) can be easily accommodated before entrance focal plane. | No space for CU accommodation at entrance FP (show stopper in the assumed de-rotation scheme) | No space for CU accommodation at entrance FP (show stopper in the assumed de-rotation scheme) |
| **Volume** | Total volume allocation is similar in the three designs, provided folding optimization is done. No detailed requirements on AOM volume allocation are available at this stage | | |

**Table 3: engineering tasks designs comparison**

**Comments to Table 3:**

[1]Total manufacturing impact in the overall error budget is similar for the different designs (Refractive and Reflective have been separately estimated, while Catadioptric is assumed to be in the same range).

[2]ROM alignment tolerances estimation has been performed taking care that the overall impact of alignment to the error budget was the same for the different designs (~15nm), assuming final focal plane focus and centering compensation. The reported values are the tightest ones.

## CONCLUSIONS

We have presented the optical design trade-off analysis of the Adaptive Optics Module of MAVIS. To restrict the parameters space, the analysis was based on some assumptions and preliminary requirements. These are either imposed by top-level requirements for the instrument, or based on experience and preliminary assessments of the required functionalities.

We considered several options spanning from purely refractive, to catadioptric and purely reflective designs. The design optimization activity led to the development of three configurations (one for every family of optical solutions) which have been compared against a set of criteria assessing their performance in terms of delivered optical quality, engineering tasks (manufacturability, alignment, volume) and cost/management/risks.

The preferred option is the use of a refractive design composed by on-axis optical elements. This configuration allows to meet the preliminary design requirements and gives enough freedom for re-optimization during the following project phases. The next step is to adapt the design based on the output of other analyses (namely AO simulations, flow down of top-level requirements, DMs trade-off analysis, update of output optical interfaces) and further detailing the optical design analysis.

## REFERENCES

[1] Arsenault, R.; Madec, P. -Y.; Vernet, E.; Hackenberg, W.; Bonaccini Calia, D.; La Penna, P.; Paufique, J.; Kuntschner, H.; Pirard, J. -F.; Sarazin, M.; Haguenauer, P.; Hubin, N.; Vera, I.; "Adaptive Optics Facility Status Report: When First Light Is Produced Rather Than Captured", The Messenger, vol. 164, p. 2-7, (2016)

[2] Francois, R. et al., *"Toward a conceptual design for MAVIS"*, this conference (2019)